\documentclass{elsart}   

\usepackage{amssymb}
\begin{document} 

\begin{frontmatter}

\title{Slowed Relaxational Dynamics Beyond the Fluctuation-Dissipation Theorem}
\author[dublin,rome]{P. De Gregorio},
\author[rome]{F. Sciortino},  \author[rome]{P. Tartaglia}, 
\author[dublin]{ E. Zaccarelli} and \author[dublin]{ K. A. Dawson} 
\address[dublin]{ 
Irish Centre for Colloid Science and Biomaterials, Department of
Chemistry, University College Dublin, Belfield, Dublin 4, Ireland}
\address[rome]{Dipartimento di Fisica  Universit\`{a} di Roma La Sapienza, 
Istituto Nazionale di Fisica della Materia,
and INFM Center for Statistical Mechanics and Complexity Piazzale Aldo
Moro 2, 00185 Roma, Italy}
 
\begin{abstract}  
\noindent
To describe the slow dynamics of a system out of equilibrium, but
close to a dynamical arrest, we generalize the ideas of previous work
to the case where time-translational invariance is broken. We
introduce a model of the dynamics that is reasonably general, and show
how all of the unknown parameters of this model may be related to the
observables or to averages of the noise. One result is a
generalisation of the Fluctuation Dissipation Theorem of type two
(FDT2), and the method is thereby freed from this constraint.
Significantly, a systematic means of implementing the theory to higher
order is outlined.  We propose the simplest possible closure of
these generalized equations, following the same type of approximations
that have been long known for the equilibrium case of Mode Coupling
Theory (MCT). Naturally, equilibrium MCT
equations are found as a limit of this generalized formalism.
%We indicate that, within the same general
%framework, it should be possible to make higher level approximations,
%leading to more general applicability.
\end{abstract}

\begin{keyword}
Fluctuation-Dissipation Theorem \sep glasses \sep Mode Coupling Theory
\sep aging
\PACS{61.20.Lc \sep 64.70.Pf \sep 05.20-y}  
\end{keyword}
\end{frontmatter}

\section*{Introduction}
Glasses are typically amorphous, dynamically slowed or arrested states
of matter, whose dynamical relaxational processes are dramatically
slower than those of liquid states, despite their great structural
similarities. It has long been known that, for many systems, slow and
careful approach to what is clearly a non-equilibrium glass transition
leads to entirely reproducible behaviour more reminiscent of phase
transitions at equilibrium. Such phenomena are often called, albeit
loosely, `equilibrium glass transitions'. The most complete current
microscopic theory available for these systems, Mode Coupling Theory
(MCT) \cite{gotze} has implicit within it \cite{lettermct} the
Fluctuation Dissipation Theorem of type two (FDT2) \cite{kubo}, but
despite this, the agreement between theory and light scattering
measurements \cite{vanmegen} are remarkably good for simple colloidal
systems. The explicit manifestation of the `equilibrium' nature of the
system is that all observables are functions only of the time
differences, and are independent of when the experimental measurements
are commenced.  However, cases such as this are by no means natural,
and typically experiments are found to depend on the time of waiting
$t_w$ after a change in the external parameters of the
system. Numerous experimental descriptions exist. However, in a
pioneering paper \cite{kurchan} on model systems it was shown that
equilibrium MCT ideas could be extended to deal with systems that
`age' and therefore violate FDT2
\cite{kob,parisi,francesco}. The ideas contained in that
paper have been helpful in developing the concept of aging, but it is
not until recently that attempts have been made to extend the existing
microscopic theory of equilibrium glassy states (MCT) in the same
spirit. Two such approaches appear to be developing. One of these is
described in this paper. The other
\cite{latz} uses an entirely different strategy by which projection
operators are extended to the non-equilibrium state.

Before beginning our discussion, we should sound some notes of
caution. Whilst the framework we build is quite general, the
approximations we use are the analogues of those used in equilibrium
MCT \cite{gotze}, and in non-microscopic theories of aging
\cite{kurchan}, \cite{models1}-\cite{models4}.  We have argued elsewhere
\cite{lettermct,longmct} that these amount to a type of dynamical 
approximation, and that is known to be accurate close to the arrest
transition only for colloidal systems. Elsewhere, as in molecular
glasses, they are approximately correct up to some characteristic
distance from the glass transition, often called the MCT temperature
\cite{emilia}. 
In terms of approximations, what we here present cannot
be expected to improve on this fundamental limitation. On the other
hand we may hope that what is described below will be the
non-equilibrium (aging) theory of those systems for which MCT
has proven of value.

\section{Fundamentals}
\noindent
It has been shown in previous works \cite{lettermct,longmct} that it
is possible to write density equations of motion starting from Newtons
equations.  This gives an explicit formula for $\ddot{\rho}_{\bf
k}(t)$ where the density variables $\rho_{\bf k}(t)$ are defined as
the Fourier transform of the number density $\rho({\bf r},t)$, i.e.
\begin{equation}   
\rho_{\bf k}(t)= \int_V d{\bf r} \rho({\bf r},t)e^{i{\bf k}\cdot{\bf r}} = 
\sum_{j=1}^N e^{i{\bf k}\cdot {\bf r}_j(t)}   
\label{rhok}  
\end{equation}

In principle, to describe the underlying Newtonian variables of
position and velocity we would need many fields, the most fundamental
of which are generally considered to be the hydrodynamic variables of
longitudinal and transverse currents, along with a local temperature
or entropy variable. Until now, mode coupling type theories have been
based on longitudinal current, and therefore only the density degree
of freedom, and we shall illustrate our points in the following within
this particular approximation. Thus, the first time derivative of
density is related to the longitudinal current, and the derivative of
this current is treated in Mode Coupling Theory (MCT), leading to an
equation involving the second order derivative of density in time
\cite{gotze}. Extensions involving more fields are more complex, 
and they will be the subject of future work.

The basic approach in any case is to develop a general model that can
describe slow out-of-equilibrium dynamics by extending FDT2
\cite{kubo}, and then to find some simple approximations to close the
model, as a starting point of applicability.
 
Thus, we write the density equations as,
\begin{equation}
\ddot{\rho}_{\bf k}(t) = {\mathcal{D}}_{\bf k}(t,\tau)+
{\mathcal{R}}_{\bf k}(t,\tau) 
\label{generaleq}
\end{equation} 
where we separate out deterministic and stochastic motions,
representing with ${\mathcal{D}}_{\bf k}(t,\tau)$ and
${\mathcal{R}}_{\bf k}(,\tau)$ respectively the deterministic and the
random force acting on the density waves.  Observables have to be
calculated by averaging over the noise distribution which has not yet
been stipulated.

We have explicitly indicated the dependence on two
distinct times in the history of the system, $t$ and
$\tau$. In general, $t$ can represent the final observation time,
while $\tau$ can be considered the so-called waiting time in an aging
experiment. 
The choice of these two times implies that, once the
variables of the system are known at time $\tau$, equations
(\ref{generaleq}) determine the dynamics of the density variables
through the unknown functions ${\mathcal{D}}_{\bf k}(t,\tau)$ and
${\mathcal{R}}_{\bf k}(t,\tau)$.  

We start by discussing some general properties that the random force
must possess. Indeed, it is fundamental to require the stochastic
process to satisfy the causality condition. Essentially all that we
will show, up to the final approximations, derive only from the form
of equations (\ref{generaleq}), and the causality requirements.  These
explicitly are,
\begin{eqnarray}&&\langle\rho_{-{\bf k}}(\tau){\mathcal{R}}_{\bf k}(t,\tau)
\rangle=0 
\label{causality1} \\
&&\langle\dot{\rho}_{-{\bf k}}(\tau){\mathcal{R}}_{\bf k}(t,\tau)\rangle=0
\label{causality2}
\end{eqnarray}
where $t$ corresponds to all times later than $\tau$.  The brackets
indicate averages over the (unknown) non-equilibrium distribution of
the system. If the decomposition into deterministic and stochastic
motions, made in (\ref{generaleq}), was exact, these averages would be
the same as the averages over the noise distribution.

Now, using (\ref{generaleq}), the two causality conditions
(\ref{causality1}) and (\ref{causality2}) can be rewritten as,
\begin{eqnarray}
&&\langle\rho_{-{\bf k}}(\tau)\ddot{\rho}_{{\bf k}}(t)\rangle=\langle
\rho_{-{\bf k}}(\tau){\mathcal{D}}_{{\bf k}}(t,\tau)\rangle  
\label{causinduced1} \\ 
&&\langle\dot{\rho}_{-{\bf k}}(\tau)\ddot{\rho}_{{\bf k}}(t)\rangle=
\langle\dot{\rho}_{-{\bf k}}(\tau){\mathcal{D}}_{{\bf k}}(t,\tau)\rangle
\label{causinduced2}
\end{eqnarray}

These also imply two further conditions, that are obtained by taking
their first derivative with respect to $\tau$, and then using
(\ref{causality1}) and (\ref{causality2}) as well as
(\ref{generaleq}). Thus, we have,
\begin{equation}
\langle\rho_{-{\bf k}}(\tau) \frac{\partial}{\partial\tau}
{\mathcal{D}}_{{\bf k}}(t,\tau)\rangle = 0 
\label{generalortho1} 
\end{equation}
\begin{eqnarray}
\langle\dot{\rho}_{-{\bf k}}(\tau)\frac{\partial}{\partial\tau}
{\mathcal{D}}_{{\bf k}}(t,\tau)\rangle &=&
\langle\ddot{\rho}_{-{\bf k}}(\tau){\mathcal{R}}_{{\bf k}}(t,\tau)\rangle 
\nonumber\\ &=&
\langle{\mathcal{D}}_{-{\bf k}}(\tau,\tau){\mathcal{R}}_{{\bf k}}(t,\tau)
\rangle+
\langle{\mathcal{R}}_{-{\bf k}}(\tau,\tau){\mathcal{R}}_{{\bf k}}(t,\tau)
\rangle 
\label{generalortho2}
\end{eqnarray}

Conditions (\ref{generalortho1},\ref{generalortho2}) constitute two of
the fundamental constraints on which to build a theory of
non-equilibrium slowed dynamics.

At this point we are still at liberty to use any trial deterministic force.
We choose to write the most general linear form, i.e.
\begin{eqnarray} 
{\mathcal{D}}_{{\bf k}}(t,\tau) &=& -\Omega_{\bf k}(t)\rho_{\bf k}(t) 
-\nu_{\bf k}(t)\dot{\rho}_{\bf k}(t)\nonumber\\
&-&\int_\tau^t dt' \gamma_{\bf k}(t,t')\dot{\rho}_{\bf k}(t') - 
\int_\tau^t dt' \chi_{\bf k}(t,t')\rho_{\bf k}(t')
\label{generalforce}
\end{eqnarray}
where the quantities $ \Omega_{\bf k}(t)$, $\nu_{\bf
k}(t)$,$\gamma_{\bf k}(t,t')$ and $ \chi_{\bf k}(t,t')$ are four
unknown parameters of the theory. We may note that so far, if one
accepts the underlying Mori-type hypothesis that dynamics may be
represented by a deterministic (slow) part, and noise (fast) part,
even to describe non-equilibrium states, the linear approximation
could be essentially exact for most conditions, providing the memory
kernels are chosen correctly.

Substituting (\ref{generalforce}) into equations (\ref{generaleq}), we
obtain the set of generalised Langevin equations for the density
variables as,
\begin{eqnarray}
&&\ddot{\rho}_{\bf k}(t) + \nu_{{\bf k}}(t)
\dot{\rho}_{{\bf k}}(t) + {\Omega}_{{\bf k}}(t)\rho_{{\bf k}}(t)\nonumber\\
&+& \int_\tau^t dt' \gamma_{{\bf k}}(t,t')\dot{\rho}_{{\bf k}}(t') + 
\int_\tau^t dt' \chi_{{\bf k}}(t,t')\rho_{{\bf k}}(t')
={\mathcal{R}}_{\bf k}(t,\tau) 
\label{langevin}
\end{eqnarray} 

With the expression (\ref{generalforce}), we assume that the
deterministic force is composed of linear terms in $\rho_{\bf k}(t)$
and $\dot{\rho}_{\bf k}(t)$, explicitly separated into instantaneous
and non-local contributions. $\Omega_{\bf k}(t)$ and $\nu_{\bf
k}(t)$ are time dependent because the system is evolving towards
equilibrium and, therefore, so are the collective
variables. $\Omega_{\bf k}(t)$ is the frequency of the elementary
excitations that would be `phonons' at equilibrium, while $\nu_{\bf
k}(t)$ takes into account the dumping of the modes. The integral terms
are memory contributions to these kinds of effects, configurational
evolution and dissipation respectively. They are not
time-translational invariant as the observables of the system are not.

It is not possible to apply a simple FDT2 as soon as this invariance
is broken. Nevertheless, the choice made in (\ref{generalforce})
enforces the width of the noise distribution to be closely related to
the memory kernels as was for the equilibrium case. This can be viewed
as a generalization of FDT2 for a system out of equilibrium. Indeed,
combining (\ref{generalortho1},\ref{generalortho2}) and
(\ref{generalforce}) we obtain,
\begin{equation}
\langle{\mathcal{R}}_{-{\bf k}}(\tau,\tau)
{\mathcal{R}}_{{\bf k}}(t,\tau)\rangle =
N\left\{J_{k}(\tau)-\frac{1}{4}\frac{\dot{S}_{k}^2(\tau)}  
{S_{k}(\tau)}\right\} \gamma_{{\bf k}}(t,\tau)
\label{gamma}
\end{equation}
and 
\begin{equation}
\chi_{{\bf k}}(t,\tau) = -\frac{1}{2}\gamma_{{\bf k}}(t,\tau)
\frac{\partial}{\partial\tau}\log{S}_{k}(\tau)
\label{chi}
\end{equation}
with the definitions,
\begin{eqnarray}    
&&S_k(t,\tau) = \frac{1}{N} \langle\rho_{-{\bf k}}(\tau) 
\rho_{{\bf k}}(t)\rangle   \\
&&J_k(t,\tau) = \frac{1}{N} \langle\dot{\rho}_{-{\bf k}}(\tau) 
\dot{\rho}_{{\bf k}}(t)\rangle \\
&&S_k(\tau) = S_k(\tau,\tau); \qquad J_k(\tau) = 
J_k(\tau,\tau)
\end{eqnarray}

Here, $S_k(\tau)$ and $J_k(\tau)$ are equal time
correlators. In the equilibrium case, clearly, they simply reduce to,
\begin{equation}
S^{(eq)}_k(t)=S_k; \qquad
J^{(eq)}_k(\tau) = \frac{k^2}{\beta m} 
\end{equation}
Also, $\dot{S}^{(eq)}_k(t)=0$.

Thus, equation (\ref{gamma}) is the generalisation of FDT2 for systems
out of equilibrium, and indeed it possesses the correct limit to the
equilibrium FDT2 \cite{kubo},
\begin{equation}
\langle{\mathcal{R}}^{(eq)}_{-{\bf k}}(\tau,\tau)
{\mathcal{R}}^{(eq)}_{{\bf k}}(t,\tau)
\rangle= \frac{Nk^2}{\beta m} \gamma^{(eq)}_{\bf k}(t-\tau)
\end{equation}
while $\chi^{(eq)}_{\bf k}(t,\tau)=0 $,
and this quantity does not appear in the equilibrium theory.  

It is still necessary to determine the generalised frequency,
$\Omega_{\bf k}(t)$, and the instantaneous part of the friction,
$\nu_{\bf k}(t)$, to characterise completely the model. Thus, we write
equations (\ref{langevin}) for the time $\tau$,
\begin{equation}
\ddot{\rho}_{\bf k}(\tau) + 
\nu_{{\bf k}}(\tau)\dot{\rho}_{{\bf k}}(\tau) + 
{\Omega}_{\bf k}(\tau)\rho_{{\bf k}}(\tau) = {\mathcal{R}}_{\bf k}(\tau,\tau)
\end{equation}
and we require the following initial conditions for our stochastic
process,
\begin{equation}
\langle\rho_{-{\bf k}}(\tau){\mathcal{R}}_{\bf k}(\tau,\tau)\rangle=0 
\label{initial1}
\end{equation}
\begin{equation}
\langle\dot{\rho}_{-{\bf k}}(\tau){\mathcal{R}}_{\bf k}(\tau,\tau)\rangle=0.
\label{initial2}
\end{equation}

These are consistent with the causality relations of equations
(\ref{causality1},\ref{causality2}), but one can imagine other models
where they are not applied. Their choice leads to the following
relations,
\begin{eqnarray}
&&\frac{1}{2}\ddot{S}_k(\tau) - J_k(\tau) +
\frac{1}{2}\nu_{\bf k}(\tau)\dot{S}_k(\tau) + 
\Omega_{\bf k}(\tau) S_k(\tau) = 0 
\label{rhoddotrho} \\
&&\frac{1}{2} \dot{J}_k(\tau) + 
\nu_{\bf k}(\tau) J_k(\tau) + 
\frac{1}{2}\Omega_{\bf k}(\tau)\dot{S}_k(\tau) = 0 
\label{dotrhoddotrho}
\end{eqnarray}
Solutions to these equations are,
\begin{eqnarray}
&&\Omega_{\bf k}(\tau) = 
\frac{4J_{k}^2(\tau)-2J_{k}(\tau)\ddot{S}_{k}(\tau) + 
\dot{J}_{k}(\tau)\dot{S}_{k}(\tau)}
{4J_{k}(t)S_{k}(\tau) - \dot{S}_{k}^2(\tau)} 
\label{Omega} \\
&&\nu_{{\bf k}}(\tau) = 
\frac{\ddot{S}_{k}(\tau)\dot{S}_{k}(\tau) - 
2\dot{J}_{k}(\tau)S_{k}(\tau) - 
2J_{k}(\tau)\dot{S}_{k}(\tau)}
{4J_{k}(\tau)S_{k}(\tau) - \dot{S}_{k}^2(\tau)} 
\label{nu}
\end{eqnarray}  

Note that the choice (\ref{generalforce}) implies four unknown
parameters, and equations
(\ref{generalortho1},\ref{generalortho2},\ref{Omega},\ref{nu}) fix
these in terms of observables, and of the noise distribution, both of
which might be considered input into the theory or determined
self-consistently. It is interesting to note that a very similar
structure has been developed by Latz \cite{latz}, using the method of
non-equilibrium projector operators.  

We shall discuss the first attempts to determine these quantities
self-consistently within certain approximations.

Following the same kind of steps made in \cite{lettermct,longmct} we
define,
\begin{equation}
\ddot{\rho}_{\bf k}(t) + \nu_{\bf k}(t) \dot{\rho}_{\bf k}(t) + 
\Omega_{\bf k}(t) \rho_{\bf k}(t) = {\mathcal{F}}_{\bf k}(t) 
\label{forceF} 
\end{equation}
where ${\mathcal F}_{\bf k}(t)$ are considered to be the couplings between
the modes. If we seek to minimize these, then the `best' choice of
modes can be determined by,
\begin{eqnarray}
&&\frac{\partial}{\partial \Omega_{\bf k}(t)}
\langle|{\mathcal{F}}_{\bf k}(t)|^2\rangle = 0 \label{derivOmega} \\
&&\frac{\partial}{\partial \nu_{\bf k}(t)}\langle|{\mathcal{F}}_{\bf k}(t)|^2
\rangle = 0  
\label{derivnu}
\end{eqnarray}
These equations give the same solutions as the ones given in
(\ref{Omega}) and (\ref{nu}), providing one means of interpreting the
choice (\ref{initial1}) and (\ref{initial2}).

Now we define the normalised correlators,
\begin{eqnarray}
\Phi_{k}(t,\tau) = 
\frac{S_{k}(t,\tau)}{S_{k}(\tau,\tau)};& & \qquad   
\Psi_{k}(t,\tau) = 
\frac{\langle\dot{\rho}_{-{\bf k}}(\tau)\rho_{\bf k}(t)\rangle}
{S_{k}(\tau,\tau)} \label{phipsi}
\end{eqnarray}   
and, using the Langevin equations (\ref{langevin}) we write,
\begin{eqnarray} 
&&\frac{\partial^2}{{\partial} t^2}
\Phi_{k}(t,\tau) + \nu_{{\bf k}}(t)
\frac{\partial}{{\partial} t}\Phi_{k}(t,\tau) + 
\Omega_{{\bf k}}(t)\Phi_{k}(t,\tau)  \nonumber \\
&&+ \int_{\tau}^{t}dt'\chi_{{\bf k}}(t,t')\Phi_{k}(\tau,t') +
\int_{\tau}^{t}dt'\gamma_{{\bf k}}(t,t')\frac{\partial}
{\partial t'}\Phi_{k}(\tau,t') = 0 
\label{phi}
\end{eqnarray}
\begin{eqnarray} 
&&\frac{\partial^2}{\partial t^2}
\Psi_{k}(t,\tau) + \nu_{{\bf k}}(t)\frac{\partial}
{\partial t}\Psi_{k}(t,\tau) + \Omega_{{\bf k}}(t) 
\Psi_{k}(t,\tau) \nonumber  \\
&& + \int_{\tau}^{t}dt'\chi_{{\bf k}}(t,t') 
\Psi_{k}(\tau,t') + \int_{\tau}^{t}dt'
\gamma_{{\bf k}}(t,t')\frac{\partial}{\partial t'}
\Psi_{k}(\tau,t') = 0 
\label{psi}
\end{eqnarray}
(\ref{phi}) and (\ref{psi}) are the equations of motion for the
system.  This set of equations is equivalent to the equations proposed
in \cite{latz}.

Now, these equations require knowledge only of the dispersion of the
noise $\langle{\mathcal R}(t,\tau){\mathcal R}(\tau,\tau)\rangle$ (see
equations (\ref{gamma},\ref{chi})), as well as $S_k(\tau)$ and
$J_k(\tau)$. 

Before turning to make useful approximations, we discuss briefly a new
method that leads to closure of these equations. It can lead to a
sequence of corrections, and, in the equilibrium case leads to the
known MCT equations within the same type of approximations.

The idea consists in applying a variational principle for that part of
the exact random force that is hard to calculate.  We shall write a
sort of constitutive relation for the noise that connects it to
density variables, but where the noise is constrained to have certain
reasonable properties, essentially ensuring that both sides of the
stochastic equations are consistent.  We therefore write,
\begin{eqnarray}
&&{\mathcal R}_{\bf k}^{VAR}(t,\tau)=
\Omega_{\bf k}(t)\rho_{\bf k}(t)+\nu_{\bf k}(t)\dot{\rho}_{\bf k}(t)
+ d^{(1)}_{\bf k}(t) \rho_{\bf k}(t)
+ \sum_{k'\neq k}d^{(2)}_{\bf k,k'}(t) \rho_{\bf k-k'}(t) \rho_{\bf k'}(t)
\nonumber\\ &&
+\sum_{k',k''\neq k'}d^{(3)}_{\bf k,k',k''}(t) 
\rho_{\bf k-k''}(t)\rho_{\bf k''-k'}(t) \rho_{\bf k'}(t) 
+ \dots \nonumber\\ &&
+\int_\tau^t dt' \gamma_{{\bf k}}(t,t')\dot{\rho}_{{\bf k}}(t') + 
\int_\tau^t dt' \chi_{{\bf k}}(t,t')\rho_{{\bf k}}(t')
\label{variational}
\end{eqnarray}
where the set of parameters $d^{(n)}$ have to be determined by making
this trial form the closest possible to the true one, that contains
$\ddot{\rho}_{\bf k}(t)$. In principle, one needs an infinite number
of the terms in the sum (\ref{variational}), and the level of the
approximation of a theory will be correspondent with the number of
terms considered.

To be acceptable this choice of noise must be `faithful', in the sense
that the Langevin process of equations (\ref{langevin}) should be
preserved.  This issue can be partially addressed by requiring that
the noise fluctuations on a single time slice are consistent with the
Langevin process (\ref{langevin}). Thus, we impose,
\begin{equation}
\frac{\partial}{\partial d^{(n)}(t)} \langle |{\mathcal R}_{\bf k}(t,\tau)-
{\mathcal R}_{\bf k}^{VAR}(t,\tau)|^2 \rangle=0
\label{fidelity}
\end{equation}
where ${\mathcal R}_{\bf k}(t,\tau)$ is defined in (\ref{langevin}).
This leads to an infinite set of coupled equal-time equations. In the
next section we shall give an example of these.

The other conditions to be satisfied by the expansion
(\ref{variational}) of the noise are the constraints
(\ref{initial1},\ref{initial2}). These conditions can now be
explicitly written as,
\begin{eqnarray}
&&\Omega_{\bf k}(\tau) S_k(\tau)+ 
\frac{1}{2}\nu_{\bf k}(\tau) \dot{S}_k(\tau)+ d^{(1)}_{\bf k}(\tau) 
S_{\bf k}(\tau)+ \frac{1}{N}\sum_{k'\neq k} d^{(2)}_{\bf k,k'}(\tau)
\langle \rho_{-{\bf k}}(\tau)\rho_{\bf k-k'}(\tau) \rho_{\bf k'}(\tau)\rangle
\nonumber \\
&&+\frac{1}{N}\sum_{k',k''\neq k'}d^{(3)}_{\bf k,k',k''}
(\tau)\langle\rho_{-{\bf k}}(\tau)
\rho_{\bf k-k''}(t)\rho_{\bf k''-k'}(\tau) \rho_{\bf k'}(\tau)\rangle
+ \dots=0
\label{constraint1}\\ &&\nonumber\\ &&\nonumber\\
&&\frac{1}{2}\Omega_{\bf k}(\tau) \dot{S}_k(\tau)+ \nu_{\bf k}(\tau)
J_k(\tau)+ \frac{1}{2}d^{(1)}_{\bf k}(\tau)\dot{S}_{\bf k}(\tau)+ 
\frac{1}{N}\sum_{k'\neq k} d^{(2)}_{\bf k,k'}(\tau)
\langle \dot{\rho}_{-{\bf k}}(\tau)\rho_{\bf k-k'}(\tau) 
\rho_{\bf k'}(\tau)\rangle \nonumber \\
&&+\frac{1}{N}\sum_{k',k''\neq k'}d^{(3)}_{\bf k,k',k''}(\tau)
\langle\dot{\rho}_{-{\bf k}}(\tau)
\rho_{\bf k-k''}(t)\rho_{\bf k''-k'}(\tau) \rho_{\bf k'}(\tau)\rangle
+ \dots=0
\label{constraint2}
\end{eqnarray}
These would be exact equations of motion determining the
non-equilibrium structure factor and currents, providing one knows the
couplings $d^{(n)}$ and the expansion (\ref{variational}) is exact.
We now discuss some simple closure ideas. These should not be regarded
as complete, but merely indicative of the sorts of approximations that
will need to be considered in future.

\section{Simplest Closure}

It is now necessary to employ some approximations regarding the
multiple averages involving densities, or equivalently the noise
distribution, as well as the kinetic terms. These last terms would be
well approximated by their equilibrium limits if we are exploring the
system after the velocities have equilibrated, and this is the line of
thinking we shall pursue as a first approximation. A different
approach has to be taken to account for the multiple averages and
also, for the equal time structure factor that is itself evolving in
time.

To make relation to the known MCT theory at equilibrium we shall
consider only the first non-linear term in the sum which represents
the noise (\ref{variational}), and thus, the only parameters left to
choose are $d^{(1)}_{\bf k}(t)$ and $d^{(2)}_{\bf k,k'}(t)$. We then
apply the minimization condition (\ref{fidelity}),
\begin{eqnarray}
&&\frac{\partial}{\partial d^{(1)}_{\bf k}(t)}
\left\langle|\ddot{\rho}_{\bf k}(t) - d^{(1)}_{\bf k}(t) \rho_{\bf k}(t) -
\sum_{k' \neq k}d^{(2)}_{\bf k, k'}(t) \rho_{\bf k-k'}(t) 
\rho_{\bf k'}(t)|^2\right\rangle=0 \label{coefficient}\\ \nonumber \\ 
\nonumber \\
&&\frac{\partial}{\partial d^{(2)}_{\bf k,p}(t)}
\left\langle|\ddot{\rho}_{\bf k}(t) - d^{(1)}_{\bf k}(t) \rho_{\bf k}(t) -
\sum_{k' \neq k}d^{(2)}_{\bf k, k'}(t) \rho_{\bf k-k'}(t) 
\rho_{\bf k'}(t)|^2\right\rangle=0
\label{coefficients}
\end{eqnarray}  

This explicitly gives the conditions,
\begin{eqnarray}
&&\sum_{k'\neq k}d^{(2)}_{\bf k,k'}(t) 
\langle \rho_{\bf -k}(t) \rho_{\bf k-k'}(t) 
\rho_{\bf k'}(t)\rangle \nonumber \\
&&+\d^{(1)}_{\bf k}(t)\langle \rho_{\bf -k}(t) 
\rho_{\bf k}(t)\rangle -\langle \rho_{\bf -k}(t) 
\ddot{\rho}_{\bf k}(t)\rangle =0 \label{coefficients1} \\ \nonumber \\ 
\nonumber \\
&&\sum_{k'\neq k}d^{(2)}_{\bf k,k'}(t) 
\langle \rho_{\bf -k+p}(t) \rho_{\bf -p}(t)\rho_{\bf k-k'}(t) 
\rho_{\bf k'}(t)\rangle \nonumber \\
&&+\d^{(1)}_{\bf k}(t)\langle \rho_{\bf -k+p}(t) 
\rho_{\bf -p}(t)\rho_{\bf k}(t)\rangle -\langle \rho_{\bf -k+p}(t) 
\rho_{\bf -p}(t)\ddot{\rho}_{\bf k}(t)\rangle =0
\label{coefficients2}
\end{eqnarray} 
to solve for finding the best possible coefficients $d^{(1)}_{\bf
k}(t)$ and $d^{(2)}_{\bf k,k'}(t)$. Equation (\ref{coefficients1}) is
trivial, as it corresponds to the same equation as
(\ref{constraint1}).

If we had chosen to work in the Newtons equation representation we
would at this point still have the bare potential present in the
problem \cite{lettermct,longmct}.  In essence, by choosing the best
form of the noise in terms of density in an independent manner, as
above, we renormalise the instantaneous forces experienced by the
density waves at a single time slice, away from what they would have
been if we had the bare potential to generate the noise correlations
\cite{lettermct,longmct}. This may also be viewed as performing a
partial sum over some of the noise to give exact, or nearly so,
equal-time averages, and leaving the averages to be calculated in the
remaining correlators to be carried out with the remaining noise. The
most important approximations are then due to approximations of the
correlations between different time slices. In a manner, we may view
this as having renormalised the vertices of the problem prior to
proceeding with any approximation of the unequal time correlators.
        
At this stage we could in principle solve for the coefficients, but
consistent with previous MCT approximations, we may further simplify
the problem.  As in equilibrium MCT \cite{gotze}, we choose a Gaussian
approximation for the four point averages and a superposition
approximation for the triplets, $\langle \rho_{-{\bf k}}(t) \rho_{\bf
p}(t) \rho_{\bf q}(t)\rangle \approx N S_k(t)S_p(t)S_q(t)\delta_{{\bf
k},{\bf p+q}}$, which consists in neglecting the triple direct
correlation function $c_3$. 

Recall that these averages are not over all the noise ensemble but
only that part remaining after the pre-averaging mentioned above. In
this sense, using the leading non-trivial approximation for averages
(Gaussian for even averages, superposition for odd averages) may not
be as severe an approximation as one would think at first sight.

Indeed, in the equilibrium case solutions of equation
(\ref{coefficients2}), in these approximations, are given by,
\begin{equation}
d^{(1) eq}_k \approx -\frac{k^2}{\beta m S_k^{(eq)}}; \qquad
d^{(2) eq}_{\bf k,k'} \approx
\frac{{\bf k}\cdot{\bf k'}}{\beta m V}c^{(eq)}_{k'}
\end{equation}
where $c^{(eq)}_{k}$ is the equilibrium direct correlation
function. We have also assumed $
\sum_{k'\neq k}d^{(2)(eq)}_{\bf k,k'} S_{\bf -k}^{(eq)} S_{\bf k-k'}^{(eq)} 
S_{\bf k'}^{(eq)} =0.$

Thus, inserting this formula in (\ref{variational}) we can calculate
(\ref{gamma}) neglecting the integral contribution, but still accepting
(\ref{phi}) as the correct equations of motions, recovering the 
well established equilibrium MCT equations \cite{gotze,longmct}. 
 
In principle, out of equilibrium, it is still possible to solve for
the coefficients $d^{(2)}_{\bf k,k'}(t)$, that minimize
(\ref{coefficients}), but to have some explicit expressions for them
we have to make further approximations than those in the equilibrium
case.  We shall, as a first approximation, consider ourselves to be in
the regime where the kinetic contributions to the observables are the
same as in the equilibrium limit, and we thereby assume that the
velocities relax towards their equilibrium values much faster than the
positions.

In the following we consider the case of any time instant $t'$, in the
range between $\tau$ and $t$.  Thus, we approximate,
\begin{eqnarray}
\langle\ddot{\rho}_{-{\bf k}}(t')\rho_{\bf k-p}(t')\rho_{\bf p}(t')\rangle 
\approx - \frac{N{\bf k}\cdot({\bf k-p})}{\beta m}S_p(t') - 
\frac{N{\bf k}\cdot{\bf p}}{\beta m}S_{|{\bf k-p}|}(t'). 
\end{eqnarray}

%We also consider that the equal time structure factor is varying in
%time only on very long scales and the contribution of $\nu_{\bf
%k}(t')\langle\dot{\rho}_{-{\bf k}}(t')\rho_{\bf k-p}(t')\rho_{\bf
%p}(t')\rangle$ can be neglected compared to the other terms in
%(\ref{coefficients2}). 
The last approximation to be made is that
$\langle\rho_{-{\bf k}}(t')\ddot{\rho}_{\bf k}(t')\rangle =
(1/2)\ddot{S}_k(t') - J_k(t') \approx -k^2/\beta m$, with
$\ddot{S}_k(t')\approx 0$.

We can now write the result for the best
coefficients in this approximations,
\begin{equation}
d^{(1)}_k (t')\approx -\frac{k^2}{\beta m S_k (t')}; \qquad
d^{(2)}_{\bf k,k'}(t')\approx \frac{{\bf k}\cdot{\bf k'}}{\beta m V}c_{k'}(t')
\label{d2}
\end{equation}
with $nc_k(t') = 1 - 1/S_k(t')$ the generalised direct correlation
function, with $n$ the number density $N/V$.  Evidently, this is a
sort of adiabatic approximation, in which, ultimately, we expect
changes in the noise distribution to arise from changes in the slowly
evolving structure.
 
We now use (\ref{langevin}) and (\ref{variational}) to obtain,
\begin{eqnarray}
{\mathcal{R}}_{\bf k}(t',\tau) &\approx& \frac{1}{\beta m V} 
\sum_{\bf k'\neq k}({\bf k}\cdot{\bf k'})c_{k'}(t')
\rho_{\bf k-k'}(t') \rho_{\bf k'}(t')+\nu_{\bf k}(t)\dot{\rho}_{\bf k}(t)
\nonumber\\
&+&\int_\tau^t dt' \gamma_{{\bf k}}(t,t')\dot{\rho}_{{\bf k}}(t') + 
\int_\tau^t dt' \chi_{{\bf k}}(t,t')\rho_{{\bf k}}(t')
\label{noiseapprox}
\end{eqnarray}
where $t'$ can be any instant between $\tau$ and $t$. The term
containing the memory kernel is neglected in equilibrium MCT
\cite{gotze,longmct}. We will, therefore, also neglect the two
integral terms, which in any case would be of higher order in our
arguments.

Now, if we approximate the four-point density correlation for
different times, by Gaussian decomposition, as before, and we neglect
the contributions arising from the term $\nu_{\bf k}(t)\dot{\rho}_{\bf
k}(t)$ in (\ref{noiseapprox}) because of higher order, we finally find
the generalised expression for the memory kernel $\gamma_{\bf
k}(t,\tau)$,
\begin{eqnarray}
&&
\gamma_{\bf k}(t,\tau)\approx \frac{4 NS_k(\tau)}
{\beta ^2m^2 V^2[4J_k(\tau)S_k(\tau)-\dot{S}_k^2(\tau)]}
\sum_{{\bf k'}\neq {\bf k}}[({\bf k}\cdot {\bf k'})^2   
c_{k'}(\tau)c_{k'}(t) \nonumber\\
&&+ 
({\bf k}\cdot {\bf k'}){\bf k}\cdot({\bf k}-{\bf k'})
c_{k'}(\tau)   
c_{|{\bf k} -{\bf k'}|}(t)]S_{|{\bf k}-{\bf k'}|}(t,\tau) S_{k'}(t,\tau).  
\label{noneqgamma}
\end{eqnarray}
and the memory kernel $\chi_{\bf k}(t,\tau)$ is then obtained
combining this expression with (\ref{chi}). 
%For consistency, in the
%present approximation scheme, the term $\dot{S}^2_k(t)$ should be
%neglected.
We can now apply successive levels of approximation to equation
(\ref{constraint1}, \ref{constraint2}) to find an equation of motion
that determines the non-equilibrium structure factor, $S_{\bf k}(t)$.
%\begin{equation}
%\dot{S}^2_k(t)=\frac{4 k^2 S_k(t)}{\beta m}\left\{1+
%\sum_{k'\neq k}\frac{{\bf k} \cdot {\bf k'}}{k^2} n c_{k'}(t) 
%S_{|{\bf k}-{\bf k'}|}(t) S_{k'}(t) S_k(t)\right\}
%\label{structure}
%\end{equation}
%where we have also used (\ref{d2}). In the same treatment, equation
%(\ref{constraint2}) results a trivial condition, since $J_k(t)\approx
%k^2/\beta m$.
In this way, we have a simple theory of the non-equilibrium dynamical
structure factor.

\section*{Conclusions}
In this paper we have derived a generalised Langevin equation that
should reasonably describe an out-of-equilibrium dynamically slowed
system. This leads to new constraints on the noise distributions
rather than traditional FDT2, but these constraints are still
practicable to apply. We then expand the noise in a `Landau type'
expansion of the density variables, leading to a non-linear
generalised Langevin process. However we conceive that some
pre-averaging of the variables has taken place so that exact equal
time averages are recovered to a high level of quality by assuming
remaining noise is only gaussian distributed. To preserve fidelity of
the Langevin process we then insist that the noise be a faithful
representation, and that the structure of the Langevin process is
still preserved. These requirements are implemented by a variational
principle to determine the renormalised coefficients of the noise, and
by implementing the orthogonality of the noise to the `slow'
variables, $\rho_{\bf k}(t)$ and $\dot{\rho}_{\bf k}(t)$.  This leads
in principle to the determination of all of the unknown coefficients
of the noise. 

Further approximations are possible but not obligatory, but they make
the contact with existing ideas easier. The outcome is that we end up
with a theory that must be considered as an `adiabatic' extension of
the existing MCT in the sense that kinetic terms are neglected,
assumed to be equilibrated. The MCT equations are therefore modified,
but the form of the memory kernel is preserved, only it now includes
the dependence on the waiting time.  In equilibrium MCT one typically
inputs any good approximation to the static structure factor for which
the YBG (Yvon-Born-Green Equation) equation of constraint should hold,
or hold approximately\cite{lettermct}. Here we deduce the
corresponding conditions, and these generate an equation of motion for
the equal time structure factor and current correlator. These
equations would be exact providing the expansion chosen for the random
force, as in (\ref{variational}), is a complete one.
%We write this
%equation in an approximate form, consistent with the one used to
%calculate the dynamical structure factor equations. 
As non-equilibrium structure factors to use as inputs are much less
known than equilibrium ones, these equations, or some similar ones, will
be crucial to close the theory.
%at least at the simplest level of approximation,
%equivalent to that of equilibrium MCT.

%We also present the equations for non-ergodicity
%factor as predicted by this new theory, showing how a new contribution
%arising from the relaxation of the non-equilibrium structure factor
%modifies them with respect to the equilibrium MCT.

The strategy we have pursued here seems different to that developed in
\cite{latz}, and it is as yet too early to see how these different
approaches will relate to each other. Probably the answer to that
question is that in the end, as with the equilibrium MCT case
\cite{longmct}, a fair degree of correspondence will emerge, though it
will be important to see if either approach leads naturally to higher
levels of approximation.

~From what we know about colloidal systems near their kinetic arrest,
this theory should be reasonably successful in describing the
phenomena qualitatively, perhaps, as with equilibrium MCT,
quantitatively. This opens the possibility to begin systematic study
of the aging of colloids and soft materials \cite{agingcolloid,aginggel}.
However, to make useful progress, in the regime where FDT2 is violated
for molecular glasses, it will require better approximations than
those shown here, and this is a matter to which we, and others, most
certainly will direct our attentions.

\section*{Acknowledgements}
The authors wish to note the most interesting discussions with
A. Latz, for which they are most grateful. They also acknowledge very
useful remarks made by W. G\"{o}tze, and discussions with K. Kawasaki,
A. Crisanti and G. Parisi. F.S. and P.T. are supported by INFM-PRA-HOP
and MURST-COFIN2000. Both Dublin and Rome groups are supported by COST
P1.

%\end{multicols}

\end{document}